# Intelligent Nanophotonics: Merging Photonics and Artificial Intelligence at the Nanoscale

Kan Yao[1,2], Rohit Unni[2] and Yuebing Zheng[1,2,*]

[1]Department of Mechanical Engineering, The University of Texas at Austin, Austin, Texas 78712, USA

[2]Texas Materials Institute, The University of Texas at Austin, Austin, Texas 78712, USA

*Corresponding author: zheng@austin.utexas.edu

**Abstract:** Nanophotonics has been an active research field over the past two decades, triggered by the rising interests in exploring new physics and technologies with light at the nanoscale. As the demands of performance and integration level keep increasing, the design and optimization of nanophotonic devices become computationally expensive and time-inefficient. Advanced computational methods and artificial intelligence, especially its subfield of machine learning, have led to revolutionary development in many applications, such as web searches, computer vision, and speech/image recognition. The complex models and algorithms help to exploit the enormous parameter space in a highly efficient way. In this review, we summarize the recent advances on the emerging field where nanophotonics and machine learning blend. We provide an overview of different computational methods, with the focus on deep learning, for the nanophotonic inverse design. The implementation of deep neural networks with photonic platforms is also discussed. This review aims at sketching an illustration of the nanophotonic design with machine learning and giving a perspective on the future tasks.

**Keywords:** deep learning; (nano)photonic neural networks; inverse design; optimization.

# 1. Introduction

Nanophotonics studies light and its interactions with matters at the nanoscale [1]. Over the past decades, it has received rapidly growing interest and become an active research field that involves both fundamental studies and numerous applications [2,3]. Nanophotonics comprises several subdomains, including photonic crystals (PhCs) [4], plasmonics [5], metamaterials/metasurfaces [6-8], and other structured materials that can perform photonic functionalities [9]. Despite the different underlying mechanisms, configurations, materials and so forth, traditionally, the design of nanophotonic devices relies on physics-inspired methods. Human knowledge, such as the physical insights revealed by the study of simple systems, the experience obtained from previous practice, and the intuitive reasoning, provide guidelines to the design process. For example, knowing that an elongated nanoparticle responds more strongly to the incident electric fields that are polarized along its long axis, and that a ring-like structure supports magnetic resonances if the incident magnetic fields are perpendicular to the plane on which the ring lies, the combination of metallic wires and split ring resonators (SRRs) was proposed to demonstrate the negative index of refraction [10], a milestone of metamaterials. The initial designs are usually examined by simulations solving the Maxwell's equations, but they are less likely to match the desired performance directly. Therefore, adjustments to a handful of parameters and re-evaluation by simulations need to be conducted repeatedly to approach the target. While remarkable success has been accomplished using this scheme, the trial-and-error procedure becomes computationally costly and time-inefficient due to the continuously increasing complexity of the nanophotonic devices.

Inverse design tackles the design task in a different manner [11]. Without the need of physical principles for the initial guess, intended photonic functionalities are obtained by optimization in the design parameter space, which, based on advanced algorithms and combined simulations, seeks a solution that minimizes (or maximizes) an objective or fitness function related to the target. In relation to solving the direct problems, optimization-based methods require comparable computation power and time. Nevertheless, they allow one to search in the full parameter space and find designs that are non-intuitive but with optimal performance.

The recent blossoming of artificial intelligence (AI), especially the subfield of machine learning has revolutionized many realms of science and engineering, such as computer vision [12], speech recognition [13], and strategy making [14], etc. Inspired by the biological neural networks, artificial neural networks have dramatically changed the paradigm of data processing and powered the development of algorithms that can "learn" from data and perform functionalities to complete complex tasks [15]. The associated technique of deep learning is thus considered a promising candidate for the inverse design of new materials [16-18], drugs [19], and nanophotonic devices [20-22] (Figure 1). In general, the role of deep learning in nanophotonic design is also to search the parameter space for a best fit of the target. But unlike optimization-based methods doing this for every task, which makes simulations recurrent efforts, deep learning algorithms are able to navigate in a smarter way by learning from a large dataset so that a solution can be found almost instantaneously after the learning phase. Without loss of design flexibility, this data-driven scheme markedly shortens the overall computation time when a common database is available for a group of applications. On the other hand, nanophotonic circuits that process coherent light are naturally suitable to build systems compatible with the framework of neural networks [23], while the speed and energy efficiency can be much higher than those of their electronic counterparts. Therefore, the application between deep learning and nanophotonics is not one-way but interactive. As their blending is just beginning, it will be timely and beneficial to present an overview on this emerging field, from which interested readers can get a general idea and determine the directions of future research [24]. We notice that some related techniques, such as topology optimization [25], inverse design [11], neuromorphic photonics [26-29], and reservoir computing [30] have been discussed by several recent review articles. Thus, we would also direct the readers to these references if interested.

The present manuscript is organized as follows. Section 2 summarizes the recent progress in nanophotonic inverse design based on optimization. Popular techniques as well as representative examples will be introduced. In section 3, we start by explaining the concept of deep neural networks (DNNs). Important applications in designing novel devices, discovering new phenomena, and revealing underlying mechanisms are then discussed with details. Section 4 is dedicated to the efforts to perform deep learning with nanophotonic circuits and optical materials as hardware. Experimental results and theoretical models for all-optical deep learning makes this topic extremely attractive and promising. Finally, concluding remarks and outlook will be given in section 5.

## 2. Nanophotonic Design Based on Optimization Techniques

Computation-wise there are many different ways to solve a photonic design problem, either direct or inverse, whereas the basis of any design strategy is that the optical properties of a given structure can be modelled with enough accuracy. For this purpose, a variety of computational tools have been developed, such as the finite-difference time-domain (FDTD) method, finite element method (FEM), boundary element method (BEM), discrete dipole approximation (DDA), and rigorous coupled wave analysis (RCWA), etc. Despite their own pros and cons in fitting different applications, these approaches solve the governing equations of light waves, i.e., Maxwell's equations. The simulated results are evaluated by the designer or an algorithm for optimization, and the updated structure is sent back to the solver for the next cycle of simulation and optimization until the specified performance is reached.

To date, most popular algorithms used in the inverse design can be categorized into two groups: the evolutionary method, such as genetic algorithm [31,32] and particle swarming optimization [33-35], and the gradient-based method, including topology optimization [36,37], steepest descent, and so forth. Other approaches based on heuristics (simulated annealing [38,39]) or nonlinear search have also been used. The main advantage of using these techniques over the traditional physics-inspired scheme is that it opens up the full parameter space and many non-intuitive designs can be obtained with optimal performance. In this section, we summarize the recent advances on nanophotonic design based on computational methods, primarily optimization techniques. Due to the complex intersections of the many applications and algorithms, priority will be given to the similarity between applications when the selected examples are grouped, while different design methods will be introduced the first time they appear in the text unless otherwise specified.

The earliest application of computational methods in nanophotonic inverse design dates back to the late 1990s, with the attempts to optimize the performance of dielectric waveguides [40] and to engineer the bandgaps of PhCs [41]. Since then, continuous progress has been made along these lines [42-53], and some previously unattainable functionalities have been made possible by using advanced algorithms and hardware [54-72]. Among the pioneers who introduced various computational techniques into nanophotonic inverse design, Sigmund and coworkers conducted a systematic study using the tool of topology optimization, which was originally developed for structural design [36,37] but has been applied to many other applications [25]. In topology optimization, the entire design domain is discretized into pixels, each being a design variable that represents the material property at that point. The total number of variables can thus be very large for a complex design task, and the structures are not restricted to any certain class of geometries. The iterative optimization procedure consists of repeated simulations and updates of the material distribution based on gradient computation. The latter is essential; otherwise the efficiency decreases dramatically, given the many design variables in topology optimization [73]. Figure 2(A) exemplifies this technique with a PhC Z-bend [74]. It is well-known that sharp bends in a waveguide will cause significant bending loss and poor transmission. Conventional optimization methods that are not free from geometric constraints solve the problem by adjusting the hole sizes and disturbing the lattice in the whole bending area. Topology optimization, in contrast, is shown to find with higher efficiency an optimized solution with only five holes being reshaped on the outer part of each bend. Despite the slight discrepancies between the designed pattern and fabricated structure, nearly 10 dB higher transmission was experimentally achieved for a bandwidth of over 200 nm. In this specific problem, the broadband property was obtained by optimizations at a single wavelength. Nevertheless, any number of wavelengths can be used simultaneously to fit any desired spectra. In addition to waveguide bends, devices with increasing complexity, such as mode converters [75] and beam splitters [55,63], have been reported by the same group.

Figures 2(B) and 2(C) show two prototypes of beam splitters designed by different approaches. In the first example, Piggott *et al*. demonstrated multi-channel wavelength splitting [76]. The specifications of this design task are the conversion efficiencies between the input and output modes at discrete wavelengths, and two different methods were employed sequentially to find the solution. At the starting point, an "objective first" strategy was adopted to take an initial guess of the structure [77], which first constrained the mode profiles to satisfy the target performance but allowed Maxwell's equations to be violated, and then minimized this violation with an optimization algorithm. Next, for fine tuning of the structure, the steepest (gradient) descent method was applied by computing the gradient of the performance metric to find its local minimum [78]. This process was under the constraint of Maxwell's equations while the permittivity was still allowed to vary continuously. The resulting layout was a complex gradient-index (GRIN) pattern with the refractive index ranging from 1 ($n_{air}$) to 3.49 ($n_{Si}$). After converting this pattern to a binary level-set representation [79], by which the material at each position can only be air or silicon, the design was optimized again using the steepest descent method for performance and bandwidth optimization around 1300 and 1550 nm wavelengths. The whole design process took ~36 hours using a graphics processing unit (GPU) accelerated FDTD solver. The final design is shown in Figure 2(B). As can be seen, the functional region contains voids of irregular shapes, but waveguide modes at the target wavelengths are routed to different output ports with low insertion losses as they propagate through the device.

Another design of nanophotonic beam splitters was showcased by Shen *et al*. [80] In this example, an unpolarized input mode is split to transverse magnetic (TM) and transverse electric (TE) components that exit the device at two different output ports. The design was based on a direct binary search (DBS) algorithm, which differs from the gradient-based methods. In brief, an area of 2.4 × 2.4 µm$^2$ was first discretized into 20 × 20 pixels. Each pixel has a size of 120 × 120 nm$^2$ and represents a silicon pillar or a void, denoted by state 1 or 0. The thickness of the device was also discretized with a step of 10 nm. A figure of merit (FOM) for optimization was then defined as the average transmission efficiency for the TM and TE modes. Following a random sequence, the state of the pixels was switched and FOM was calculated. If FOM was improved after a switch, that pixel would retain the new state; otherwise it kept the original value. After all the pixels were addressed, a similar optimization was applied to the device thickness by changing its value to the adjacent states (±10 nm). Walking through the 400 pixels and making a slight adjustment to the device thickness completed one iteration. The optimization was terminated when the improvement of FOM was smaller than a threshold after an iteration or the maximum iteration number was reached, which took ~140 hours. As a non-gradient approach, the DBS is computationally intensive and becomes less efficient when the number of design variables increases [73]. To maintain the computation time within an acceptable range, parallelizing the algorithm and using larger clusters of processors would be necessary [81]. Figure 2(C) reports the design, experimental and simulation results, showing reasonably good agreement. On-chip devices form a class of applications suitable for inverse design. Besides beam splitters, optical diodes that perform asymmetric spatial mode conversion [70,71] (Figure 2(D)) and reflectors [82] (Figure 2(E)) have also been demonstrated using different algorithms.

Flat optics is another class of applications powered by optimization methods [7,8]. Having resonant elements arranged at an interface or in a few layers to carry out functionalities, metasurfaces and metalenses contain many variables to be carefully determined in the design process, and the problem could be larger-scale if the structure is aperiodic. For the design of dielectric metasurfaces, topology optimization could be a well-suited option [25]. Starting from a random, continuous refractive index distribution bounded by the indices of air and the dielectric ($n_d$), structures satisfying the desired performance but with binary indices 1 and $n_d$ can be achieved by using gradient-based optimization algorithms. Sell *et al*. reported an approach for designing periodic silicon metasurfaces with multiwavelength functionalities [83-86]. When the target was set to deflect light of discrete wavelengths to different diffraction orders, an FOM involving all the diffraction efficiencies was defined for optimization. By using an adjoint-based method [87], i.e., solving the forward diffraction problem and an adjoint problem that reverses the incidence to the target diffraction order directions as one iteration, the gradient of FOM can be calculated, based on which the refractive indices at each point received a small adjustment towards $n_{air}$ or $n_{Si}$. The iteration continued until a binary profile was obtained, as shown in Figure 3(A). Similar procedures can be used to tackle aperiodic and multilayer structures and to achieve different functionalities, such as wavefront manipulation, polarization control, and beam shaping. Although considerable efforts have been devoted to alleviating the aberration in metasurfaces, success was only attained in eliminating the chromatic aberration [88]. The suppression of angular and off-axis aberrations in single-layer metasurfaces is fundamentally impossible, whereas the design of multi-layer structures with an angle-dependent phase profile is practically challenging for traditional approaches. Taking advantage of topology optimization, Lin *et al*. designed a two-dimensional (2D) metalens that is free of angular aberration [89]. Figure 3(B) depicts the layout, which is symmetric but aperiodic, comprising five layers of silicon gratings embedded in an alumina background. FDTD simulations revealed that at the three target off-axis angles of incidence, light was all focused on the focal plane following the identical diffraction limit, as shown in Figure 3(C).

Evolutionary algorithms are also widely used in metasurface design. The general strategy is to maximize a fitness function by repeatedly evolving a population of candidate solutions with sequential application of selection, crossover, and mutation, etc. One possible framework consisting of four steps is illustrated in Figure 4(A), which was adapted by Huntington *et al*. for optimizing a lattice of circular holes in a metal film to achieve unique focal properties [90]. The design began with the generation of the initial population (step 0), a group of 600 randomly created binary patterns. Because the lattice consists of $33 \times 33$ holes, each denoted as 1/0 when the hole is open/closed, in total there are $2^{1089}$ possible arrangements. The field distribution for an arbitrary profile can be calculated by adding up the pre-stored complex fields from each individual hole. Compared with simulating the entire structure, this scheme significantly increases the efficiency.

For a specification of the focal behavior, a fitness function can be defined accordingly, which is maximized when the far-field intensity satisfies the target functionality [91]. In step 1, each member in the initial population was evaluated by the fitness function. The population was then sorted in step 2 by fitness, and the best-fit individuals were selected in step 3 to create a new generation of the population through a combination of crossover and mutation. The design cycle continued until a convergence condition was reached (step 4). In Ref. [90], after optimizing the parameters of the algorithm, e.g., the population size and mutation rate, a specified design task can be finished in ~210 generations within 30 minutes. But in general, since the optical properties of each individual structure is simulated rather than pre-stored, the design would take a much longer time. In fact, the high computational cost for large-scale design is an important limitation of evolutionary methods. Figure 4(B) shows a lattice which exhibits five focal points arranged in a T-shape. In addition to the 1/0 state, the holes were further encoded by their sizes. Although the same pattern can be produced with a fixed hole size, the design with three different hole sizes improved the diffraction efficiency from 55% to ~74%. Not only the far-field patterns but also the near-field intensity can be engineered using this technique. Fitchtner *et al*. studied the near-field enhancement in a checkerboard-type structure [92], which was optimized using an evolutionary algorithm. The results led to a novel "matrix nanoantenna" structure that can provide a two-fold near-field enhancement compared with a dipole antenna. Analysis of a reduced model in Figure 4(C), which retains the important structural features of the fittest design, revealed that the enhancement is caused by the complex interplay between a fundamental split-ring mode and a dipole mode in the two arms. Moreover, it is shown that by connecting nanobars to the ends of an SRR, the fundamental resonance can be shifted from near-infrared into the visible regime. Tuning the resonances of nanoantennas further enables color generation [93]. With the assistance of evolutionary algorithms combined with an electrodynamic solver based on the Green's dyadic function, Wiecha *et al*. demonstrated polarization-dependent color pixels [94]. Figure 4(D) shows the gallery of the designed silicon nanoantennas, which were optimized to have maximized scattering at 550 nm for incident polarization along the *x*-axis and at various wavelengths for incident polarization along the *y*-axis. Each pixel may have multiple elements and interestingly, as the target wavelength increases, these elements tend to merge together. The polarization-filtered dark-field spectra and images for orthogonal polarizations are compared in Figure 4(E), showing reasonable agreement with the simulated results. The design methods can also be combined with fabrications. Lee *et al*. integrated the processing steps of wrinkle lithography with the concurrent design procedure of quasi-random light-trapping nanostructures for absorption enhancement [95]. Specifically, the processing patterns were represented statistically by the Fourier spectral density functions [96], which used only three variables to connect the structure and the optical property, making the problem solvable for a genetic algorithm. Figure 4(F) depicts the optimization history of the averaged absorption enhancement by the designed structures. After ~150 iterations the search gradually converged, resulting in an enhancement of 4.7 over the weakly absorbing interval of amorphous silicon from 800 to 1200 nm.

The suppression of light scattering by an object is a topic of broad interest. In recent years, progress has been made with both forward design methods, such as transformation optics [97-101] and scattering cancellation [102], and inverse design approaches [103]. Genetic algorithms are usually adopted for optimizing a multilayer particle or cylinder to achieve omnidirectional scattering reduction [104,105], while topology optimization is more practically associated with designing bidirectional cloaks or resonators that can be realized by a low-index material [106-112], although in theory it can work for any imaginable objective function [106,113]. Figure 5 summarizes a few examples based on topology optimization. Since the design methodology does not differ much from the examples above, we will not proceed further to the details.

Lastly, we briefly outline a few other computational methods and applications. In addition to device design, optimization algorithms have been used to explore new physics. In the study of optical tweezers and optical manipulation [114-118], traditionally attention was paid to the optimization of particle geometry and multiplexed optical traps. Lee *et al*. applied constrained optimization [119], a derivative-free algorithm, together with a BEM solver to maximize the optical torque on a gold nanotriangle [120] (Figure 6(A)). At the dipole and quadrupole resonance wavelengths of the particle, a large portion of 2000 random initial illumination conditions resulted in over 5-fold enhancement of optical torque per intensity, compared with that from a standard circularly polarized planewave incidence. The optimal design at the quadrupole resonance could even lead to a 20-fold improvement, as revealed in Figure 6(B). This result provides new insights into the optical manipulation of objects with structured light and the computational framework can be generalized to opto-mechanical applications. Lin *et al*. demonstrated, based on topology optimization, that the third-order Dirac points formed by the accidental degeneracy of modes belonging to three different symmetry representations can be realized in inverse-designed PhCs

[121] (Figure 6(C)). Moreover, the third-order exceptional points (EP3) can be created by introducing a small loss term, giving rising to strong modifications in the local density of states (LDOS) and potential connections to topological photonics [122]. Topology optimization is not the only technique compatible with the binary representation of materials. Other algorithms, such as simulated annealing and particle swarm optimization, have also been used in the optimization of nanostructures. Simulated annealing mimics the process of heating and controlled cooling of a solid for recrystallization in metallurgy [39]. At each iteration of the search, the algorithm keeps every better solution, and, by choosing a temperature-dependent acceptance function, allows with a slowly decreasing probability some worse solutions to stay in the pool. Therefore, this strategy largely avoids being trapped in local minima and statistically guarantees finding a good solution, but meanwhile, its efficiency is lower than gradient-based methods. Figure 6(D) shows the design of a binary plasmonic structure composed of pixelated grooves [123]. Complex interference patterns of surface plasmon polaritons (SPPs) can be generated, showing the potential as plasmonic couplers. Particle swarm optimization works based on the movements of a population of candidate solutions (particles) in the search space. During optimization, the initially randomly distributed particles continue moving towards the then-current optimum particle in the swarm, until a certain termination criterion is reached [33]. Figure 6(E) reports the imaging of subwavelength holes by a binary super-oscillatory lens [123], which was fabricated on a 100-nm-thick aluminum film on glass and mounted to a microscope lens. The structure creates a delicate balance of the interference of a large number of diffracted beams, ensuring the hotspot is very sensitive to the presence of small objects. In addition to imaging, other reported applications include field enhancement engineering [125], waveguide design [126], and color filters [127], etc.

## 3. Nanophotonics Enabled by Deep Learning

Deep learning, as a subfield of machine learning and AI, has attracted increasing attention due to its great success in computer vision [12] and speech recognition [13] and its astonishing progress in various applications such as strategy making [14]. Recently, owing to its extraordinary capability in finding solutions from an enormous parameter space, researchers have started using deep learning in drug discovery [19], materials design [16,18,128], microscopy and spectroscopy [129-134], and other physics-related domains [135-137]. Among all these attempts, nanophotonics turns out to be a unique field, because it not only benefits from deep learning for the inverse design of advanced devices and performance improvement of existing techniques, but can also give feedback, providing platforms to implement deep learning algorithms that can operate at the speed of light and with low energy consumption. In the following, we discuss how the field of nanophotonics is actively interacting with the emerging technique of deep learning. Specifically, the recent advances in applying deep learning for nanophotonic design will be reviewed in this section, and the optical implementations of neural networks are left to the next section.

We start with a brief introduction on some basic concepts about deep neural networks (DNNs). Figure 7 illustrates the typical architecture of a DNN consisting of multiple processing layers, including an input layer at the bottom, an output layer on the top, and at least one hidden layer (but usually more) in between. Each circle in the diagram represents an artificial neuron, which is connected to other neurons in the neighboring layers by different weight values subject to learning. The input and output layers both have a fixed number of neurons or units, determined by the size of the feeding data (for the input layer) and the task of the DNN (for the output layer). In nanophotonic applications, they could correspond respectively to, e.g., the design parameters of a nanophotonic structure and its optical properties or vice versa. The hidden layers establish a nonlinear mapping between the input and output via training, from which very abstract relationships can be discovered to make predictions on the optical properties of given nanostructures and determine the design parameters for desired performance.

What lies at the very root of DNNs is the organization of the neurons. Specifically, enabled by the use of the backpropagation algorithm, a unique feature of DNNs is that data can be transformed bidirectionally through the network between the input and output layers. Conducting deep learning is thus divided into two processes, the forward inference and the training based on backpropagation, as sketched in Figure 7. In general, the computation of DNNs is achieved by matrix multiplications. In the forward inference, starting from the input layer, the neurons carrying input data form a vector $X$. Under full connectivity, each neuron $x_i$ in $X$ is connected to each neuron $y_j$ in the first hidden layer, $Y^1$, by a weight $w_{ij}$. An initial value $z_j = \sum w_{ij} \cdot x_i$ is given as a weighted combination of the neuron values from the previous layer. It is then essential to apply a nonlinear activation function $f$ to $z_j$; its importance will be explained shortly later. Now the value of neurons in the next layer $Y^1$ is rectified to $y_j = f(\sum w_{ij} \cdot x_i)$ or simply $Y^1 = f(W \cdot X)$, an expression of matrix multiplication. For the forward inference, each next layer $Y^{l+1}$ is connected to the previous layer $Y^l$ by a similar weight matrix, and this operation is repeated for all layers until arriving at the output layer $Y^L$ for a network with $L$ layers, which gives the first guess of the target $t_k$.

Because the initial values of the weights $w_{ij}$ are usually randomly chosen or at least not well suited for the problem to be solved, this guess is very likely far away from the correct answer, giving a large error. The training process works in the reverse direction. Based on the error at each output neuron, a cost function $C$ can be defined, which is minimized when the result of the forward inference, $Y^L$, is equal to the real answer. The gradient of the cost function can be calculated by the partial derivative of the cost with respect to each weight variable, $\partial C/\partial w_{ij}$. Using the chain rule of derivatives, this can be expressed by $\partial C/\partial w_{ij} = (\partial z_j/\partial w_{ij})(\partial y_j/\partial z_j)(\partial C/\partial y_j)$. When calculating the final term for the weights connected to the output layer, it is simple as the cost is directly a function of the values at those neurons. When doing the same for previous layers, the derivative of the cost with respect to each neuron's value is a weighted sum of multiple errors, because the neuron is connected by multiple routes to all the neurons in the output layer. Hence, the backpropagation is meant to efficiently calculate the partial derivatives and how the error propagates through each layer. Finally, each weight is adjusted by the partial derivative of the cost with respect to that weight, further scaled by a factor $\eta$ called the learning rate, $\Delta w = -\eta \cdot \partial C/\partial w$. Sometimes a stochastic factor is also included. This weight update is how the network learns. The process of feeding data in, calculating a prediction through the forward inference, calculating the cost by comparing it to the true values of that training data, and calculating the gradient of the cost and adjusting each weight value is repeated for many times. With a sufficiently large amount of training data, the performance of the DNN can be continually improved.

While the basic principle of deep learning is summarized above, the actual implementation of DNNs is much more complicated and contains many subtle problems, such as the choice of training data, the cost function, network depth, initial weights, and the learning rate, etc. Since these detailed techniques are not among the focus of this article, interested readers may refer to some latest topical reviews or books, e.g. Ref. [138]. Before proceeding to the applications in nanophotonic design, two concepts that will appear in the later discussion deserve a glance. First, the nonlinear function $f$, or termed as activation function or transfer function, plays an important role in DNNs. Compared with linear functions, nonlinearity allows a network to tackle more abstract representation and learn much faster with fewer neurons. Popular choices of the nonlinear function include the logistic functions, the hyperbolic tangents, and the rectified linear units

(ReLUs), etc. Second, the organization of neurons varies. Besides the fully connected network in Figure 7, another widely used architecture is the convolutional neural network (CNN or ConvNet). In such structures, data flow in the form of multiple planes, and it is a filter or kernel consisting of a small array of weights that connects the input and output planes. With this change, CNNs contain much fewer connections than standard models with a similar depth and are thus easier to train, while their theoretical best performance only decreases slightly [12]. This characteristic is highly desirable when processing high dimensional data, such as images and videos.

Applying deep learning algorithms to the nanophotonic inverse design introduces remarkable design flexibility that can go far beyond that of conventional methods based on an intuitive initial guess and many cycles of trial-and-error modeling, fabrication, and characterization. It also enables, without recurrent efforts in conducting time-consuming simulations, fast prediction of complex optical properties of nanostructures with irregular shapes and intricate architectures. A bidirectional DNN that can achieve both the design and characterization of plasmonic metasurfaces was first reported by Malkiel and coworkers [20]. For the implementation, two standard DNNs were used to perform the inverse design and spectra prediction tasks for arrays of "H"-shaped gold nanostructures. Instead of training two networks separately and composing them afterwards, it is shown that combining the networks during training is more effective and helps to avoid unstable processing. The full structure of the combined network is shown in Figure 8(A). A geometry-predicting-network (GPN) is used to solve the inverse problem, for which the training data comprise two spectra for orthogonal linear polarization excitations and the dispersive material properties. These three groups of data were fed separately and in parallel into three DNNs before they join a larger fully connected DNN. This architecture allows better representation of each data group and results in better performance if the depths of the networks are properly selected. The output of GPN includes eight design parameters, corresponding to the length, width, orientation, and existence of the five elements (four arms and one connecting bar) of a general "H"-shaped particle. The second part works on top of the GPN and functions as a spectrum-predicting-network (SPN), which receives the predicted design parameters, material properties as well as a polarization indicator as an input and returns the predicted transmission spectra as the outputs. Due to the two-phase structure of the network, backpropagation is optimized between the GPN and SPN for stability and efficiency. With a training dataset containing over 16500 geometries simulated by an FEM solver, the desired performance was achieved. Note that although the generation of training data is still a time-consuming process, the training takes only ~2 hours to get the best results. More importantly, these efforts are nonrecurrent. Once the training is complete, a query to the DNN about either the design (for a pair of given spectra) or the spectrum prediction (for a given geometry) can be solved in a few milliseconds. In contrast, the same query to an evolutionary algorithm or other traditional optimization methods would take much longer time to search the entire parameter space. Figure 8(B) shows the representative results for two testing samples, which were fabricated, measured, and composed of geometries not used in training. Excellent agreement was achieved between the retrieved parameters and real dimensions measured by scanning electron microscopy (SEM), and the spectra from measurements, predictions, and simulations based on retrieved geometries also show fairly good overlaps.

The above framework optimizes a few parameters that describes a certain form of geometries. While it shows the potential as a powerful tool of inverse design, in many circumstances varying the geometry within a single class of topology is insufficient to generate the intended complex optical responses. Liu *et al*. proposed an alternative method to explore the enormous design space by employing a generative model [22], as shown in Figure 8(C). The full architecture is constituted by three parts, namely the generator, the simulator, and the critic, all being CNNs. Specifically, the generator and critic together function as a generative adversarial network (GAN), which, unlike the previous case relying on paired input-output training data, is an unsupervised learning system [139]. In practice, the simulator was pre-trained with 6500 full-wave FEM simulations for a broad variety of shapes of gold nanoparticles. After training, its weights were fixed, and the transmittance spectra of any input patterns would be approximated by the simulator instead of being computed by full-wave simulations. The function of the generator is to create unit cell patterns of the metasurface for input spectra $T$ such that when the generated patterns are fed to the simulator, the approximated spectra $T'$ would largely replicate the original inputs $T$. However, if there is no constraint on the training, the generated patterns can be arbitrary and include numerous unrealistic results. On the other hand, if the true patterns corresponding to $T$ are directly used to determine the cost function, the model reduces to the supervised case and the pattern generation becomes a deterministic problem. Therefore, the critic plays a key role in the GAN. In the training process, it receives as inputs both the original patterns resulting in $T$ and the generated patterns in the form of images. The critic compares the two sets of images and restricts the generator to create patterns that share common features with the original structures but are not identical to them. Figures 8(D) and 8(E) summarize two examples to show the performance of the network, which was trained with several classes of geometries, such as circles, arcs, ellipses, crosses, handwritten digits, and so forth. In Figure 8(D), the network responded correctly to the query of replicating the transmittance spectra of an elliptical nanoparticle array. The generated pattern and resultant spectra (right panel) only exhibit slight deviations compared with the inputs (left panel). This retrieval can be achieved not only when the critic was fed with a single class of geometry but also for a mixed training dataset that contains all the classes, meaning that the GAN can identify the correct topology and conduct inverse design effectively. In Figure 8(E), after the critic was trained with an incomplete set of handwritten digits, the network was asked to replicate the spectra of a metasurface consisting of the missing digit "5". Very interestingly, the generated pattern was a modified "3", which departs from the ground-truth but also contains some similar geometric properties to reproduce most of the spectral features.

The fact that DNNs can learn complex functions from abstract data representations provides unprecedented opportunities to the nanophotonic inverse design. In many applications, the relations between desired functionalities and the design parameters are very intricate, and physical insights and intuitive reasoning may not help to guide the design process. One example is the generation of optical chiral fields. Chirality is a structural property of objects. An object is chiral if it cannot be superimposed to its mirror image. Due to its universal existence in nature, chirality has aroused enormous research interests [140-144]. From a nanophotonic point of view, designing chiral nanostructures that respond to chiral light (usually left/right circularly polarized (LCP/RCP) light) differently is of both fundamental and application-wise importance. However, due to the complex, unrevealed underlying mechanism, despite some requirements on symmetry being formulated to judge whether a structure is chiral or not [145,146], there is no general guidelines that can be referred to if one wants to design a structure for a given chiral response or how this response will evolve when the structure transforms. Taking advantage of deep learning, Ma and coworkers demonstrated a purpose-designed learning architecture for implementing on-demand design of three-dimensional (3D) chiral metamaterials [21]. Figure 9(A) illustrates the general form of the unit cell, which consists of two layers of gold SRRs atop an optically thick gold reflector. The two SRRs are sized $l_1$ and $l_2$ respectively and are twisted for an angle $\alpha$. These parameters together with the spacer thicknesses $t_1$ and $t_2$ define the structure of the chiral metamaterial. Upon illumination of LCP and RCP light, the metamaterial absorbs (or equivalently, reflects) the incidence differently, resulting in circular dichroism (CD) signals. Like in the previous cases, training a DNN to establish nonlinear mappings between the design parameters and spectra is possible. And indeed, when the reflection spectra were fed into a bidirectional DNN, denoted as the primary network (PN) in Figure 9(B), both spectra prediction and design retrieval can be achieved after learning from 25000 simulated samples. Note that compared with the fully connected DNN used in Ref. [20], here a tensor module is introduced to account for the size mismatch between the low dimensional input vector of five design parameters and the high dimensional output of spectra [147]. In Figure 9(C) (top panel), the dashed and solid curves show the simulated and predicted spectra respectively. However, while the two sets of spectra coincide well over most of the frequency range, obvious discrepancy occurs at the steep resonance valley near 60 THz for the co-polarization reflection of RCP. The reason of this degradation is that for each neuron in the output layer, the probability distribution generated by the nonlinear function is centered at its off-resonance value. Sharp resonance features deviating from the mean value is thus difficult to predict with high accuracy. A feasible way to fix this issue is to combine the PN with an auxiliary network (AN) that associates the design parameters directly to the CD signal (Figures 9(B) and 9(C), lower panels). Because CD is not independent of the reflections, including CD spectrum in the dataset of PN will not improve the leaning performance. When the predicted CD exceeds a threshold value, the AN triggers a fine-tuning network in the combiner to locally refine the reflection spectra near that frequency. The AN-corrected reflections are denoted by dotted lines in Figure 9(C), showing excellent agreement with the simulated results. This combined network enables on-demand inverse design of 3D chiral metamaterials and discovery of some interesting, counterintuitive phenomena. For instance, in Figure 9(D), chiral metamaterials with 10 and 170º twisting angles exhibit strong chiral responses at 60 THz when the SRRs are in proper sizes; while if following the previous argument of symmetry, they are close to the achiral structures where $\alpha = 0$ or $180º$ and are not expected to generate strong CD.

Another example of employing DNNs for inverse design without intuitive guidelines is reported by Pilozzi *et al*. In Ref. [148], they applied a deep learning algorithm to solve the inverse problem for topological photonics. Stemming from the photonic analogue of quantum anomalous Hall effect in electronics, topological photonics studies the creation of interfacial phonon transport or edge states that are protected from scattering [122]. The realization of such systems with nontrivial topological properties usually requires using magnetism, time-domain modulations or optical bianisotropy, but none of them can be easily designed for an intended edge state at a given frequency, even in the simplest one-dimensional (1D) systems. Figure 10(A) illustrates the dielectric function profile of a multilayer structure with Harper modulation [149,150], from which synthetic magnetic fields can occur. The existence of edge states and their dispersion relations can be determined by assigning proper boundary conditions and solving the eigenvectors of the transfer matrix. The band diagram for a chosen modulation profile is shown in Figure 10(B), where the green and orange strips correspond to the bandgaps, and edge states, indicated by the white crosses, exist only in the gaps where a complex function Q changes sign. Provided a modulation frequency and materials A and B, the search for layer thickness $\xi$ and the phase $\chi$ of Harper modulation for an edge state at frequency $\omega_t$ cannot be solved analytically but can be addressed by two DNNs. Because of the folding and multivalued branches of Brillouin zone, additional categorical features were included in the dataset to label the different modes and different trends of the iso-frequency surfaces. Unphysical solutions to the inverse problem were ruled out by making a self-consistency check between the predicted frequency and the ground truth in the training dataset. Figures 10(C) and 10(D) report the solutions from the direct and inverse DNNs respectively, showing good agreement with the training set (colored curves).

A keystone of inverse design via deep learning, as a data-driven method, is the sufficiently large training dataset, which is usually obtained by numerical simulations. For most of the applications above, thousands of simulations are conducted for weeks to give a good representation of the input space. To this extent, applications and geometries that are consistent with analytical methods provides a suitable playground for efficient data generation. Figures 11(A) and 11(B) exemplify this by considering the scattering problem of a $SiO_2/TiO_2$ multilayer spherical particle [151]. With analytical solutions derived from the transfer matrix method, 50000 samples were generated for different combinations of shell thickness. A fully connected DNN was used to solve the inverse design task. As can be seen in Figure 11(A), for an arbitrarily chosen sample spectrum from the test set (blue curve), the DNN successfully captures all the spectral features with only moderate deviations in amplitude at a few peaks/valleys (red dotted curve). In contrast, when a nonlinear optimization based on the interior-point method was employed to solve the same problem, much larger inconsistency is observed (black dashed curve). In fact, as the number of design parameters increases, optimization methods tend to become stuck in the local minima instead of the global ones, while DNNs are not affected. Moreover, DNNs can be easily adapted to fit different design requirements. For example, by using a different cost function, a $SiO_2$/silver multilayer particle showing broadband scattering within the desired wavelength range was found from the enormous possible candidates (Figure 11(B)). The similar procedure has also been used to study the transmission of multilayer thin films, where hundreds of thousands of samples were generated for training [152]. Since for such non-resonant structures there is a high likelihood that different configurations can result in nearly identical optical responses, a tandem network was proposed to overcome the non-uniqueness issue. As shown in Figure 11(C), a pre-trained network solving the direct problem was connected via an intermediate layer *M* to the original DNN. This architecture works in a similar way to the bidirectional networks, which applies additional constraints to the learning process. The modified network gives reasonable designs even when asked to fit some unrealistic spectra as in Figure 11(D). More complex functions, such as achieving phase delays at multiple wavelengths, can be realized by using structured thin films (Figure 11(E)) and modifying the network structure accordingly, while generating new data set is not a high cost.

Another aspect of the considerations about data is the volume. So far, the applications of deep learning in nanophotonics are limited to establishing the mappings between the design parameters and the optical responses given by spectra. The distributions of electric and magnetic fields and their derived quantities in the 2D or 3D space are also of great interest. However, using 2D or 3D field distributions as data sets is not practical. On one hand, this leads to huge amounts of data that are unaffordable for storage and training, especially in nanophotonics where ensuring high spatial resolutions is a basic need. On the other hand, how the feature representation in 2D and 3D data can be effectively utilized is largely unexplored. Barth and Becker proposed an interesting technique based on a machine learning algorithm, though not deep learning, for the classification of the photonic modes in a PhC [153] (Figure 12(A)). This task is aimed at a type of applications different from inverse design. Taking sensing for example, the 3D field distributions associated with nanostructures need to be evaluated by some criteria and optimized accordingly in order to facilitate the light-molecule interactions and maximize the performance. However, the analysis of field distributions is difficult, usually solved by visualizations and processing the full set of 3D data. Taking advantage of an algorithm based on Gaussian mixture method (GMM) [154], Ref. [153] showcased that the clustering model can reduce the 3D field distributions to a finite number of distribution prototypes, which allows the identification of characteristic photonic modes. Figure 12(B) compares the band diagrams of a PhC obtained by two methods: The left panel is calculated by an integration of electric field energy density over the volume within the hole and a thin layer above the PhC, while the right panel depicts the classification map of field distributions on the three symmetry planes marked in Figure 12(A). A fairly good match is observed, which confirms the validity of the procedure. Moreover, based on the clustering results, the field distribution prototypes can be obtained. As shown in Figure 12(C), by inspecting the field distribution of each cluster, leaky modes that result in strong near-fields can be distinguished from the radiative modes. These results were further validated by finite element simulations in Figure 12(D), where the illumination conditions were determined using the silhouette coefficients for classification [155]. Therefore, with a lower dimensional dataset, the proposed technique provides an alternative approach to extracting information from 3D field distributions that may not be accessible via visualization-based analysis.

## 4. Deep Learning on Nanophotonic Platforms

The arrival of the era of big data has rendered the speed, energy consumption, and information density of computing the key considerations in hardware development. However, after decades of continuous improvements following Moore's law, electronics started facing bottlenecks on these aspects, which are physically fundamental and can no longer be resolved by scaling. Integrated photonic circuits are considered promising candidates to overcome the above obstacles, because of the higher speed and energy efficiency associated with photons. On the other hand, traditional electronic components such as the central processing units (CPUs) are not well suited to serve the emerging techniques in artificial intelligence. New hardware architectures aimed at accelerating AI and deep learning are also in a pressing need. Within the domain of electronics, GPUs, vision processing units (VPUs), tensor processing units (TPUs) [156], TrueNorth [157] and other integrated chips [158] have been developed and showed great potential in practical applications. Meanwhile, hybrid opto-electronic systems for implementing spike processing [159,160], neuromorphic computing, and reservoir computing have also been demonstrated, and progress is being made towards their photonic realization. Some timely reviews have given comprehensive discussions on these topics [26,28,30]. In this section we keep our focus on the all-optical implementation of DNNs.

While photonic circuits in general operate at higher speeds and with higher energy efficiencies compared with their electronic counterparts, implementing DNNs and computing on photonic platforms offers a few advantages in this specific application [23,161,162]. First, as discussed in the previous section, the computation of DNNs is mostly achieved by matrix multiplications. In nanophotonic circuits, linear matrix operations can be performed very fast—almost at the speed of light, and, in parallel and efficiently due to the non-interacting nature of photons. Second, the nonlinear functions in the DNNs can be realized by optical nonlinearities in photonic circuits, such as saturable absorbers or amplifiers. Third, for a given photonic DNN, after training, the whole system is passive and consumes no power. Last, it is possible to conduct training of photonic DNNs by an optical means. This could significantly accelerate the learning process and further reduce power consumption.

One of the earliest demonstrations of photonic DNNs was reported by Shen *et al*. [23], where vowel recognition was achieved showing comparable performance to a 64-bit electronic computer. To implement the full map of a DNN after training, each layer of the network is composed of an optical interference unit (OIU) to carry out the linear matrix multiplication and an optical nonlinearity unit (ONU) that acts as the nonlinear activation. Four different vowel phonemes spoken by 90 different people were used to train and test the circuit, which were first preprocessed on a computer and then fed into the nanophotonic DNN as amplitude-encoded optical pulses to generate outputs, as shown in Figure 13(A). The physical realization of the network in a nanophotonic circuit is not as straightforward as it appears in the schematic. Because after training the weights may end up with an arbitrary distribution, the design of OIUs needs to tackle the problem of how the propagation of optical pulses through the unit can be equivalent to a multiplication by an arbitrary matrix. Fortunately, a real-valued matrix $M$ can be expressed by $M = U\sum V^\dagger$ via singular value decomposition [163], with $U$ and $V^\dagger$ denoting two unitary matrices and $\sum$ a diagonal matrix. In nanophotonic circuits, a unitary matrix can be implemented with beam splitters and phase shifters [161,164], and a diagonal matrix can be realized by using optical attenuators or amplifiers [165,166]. Therefore, a proper arrangement of these optical components is capable of performing matrix multiplications. Figure 13(B) shows the optical micrograph of an OIU fabricated on a programmable nanophotonic processor. The unit consists of 56 Mach-Zehnder interferometers (MZIs) with each of them containing two phase shifters and a directional coupler to achieve desired functionalities via programming. The red- and blue-highlighted meshes denote the components that perform the unitary and diagonal matrix multiplications, respectively. Input optical pulses propagate through the unit, producing the correct interference patterns at the output.

A similar scheme was adopted for implementing a different processor architecture for quantum transport simulations [167]. In Ref. [23], instead of physical realization, the nonlinear activation was simulated on a computer as saturable absorbers. Whether real nonlinear optical elements can work equally well is still an open question to be addressed. Figure 13(C) compares the correlation matrices of a bilayer nanophotonic DNN and a computer for the vowel recognition task. The nanophotonic DNN shows a correctness of 76.7%, lower than but still comparable to the result of 91.7% from the electronic processor. Larger error from the nanophotonic prototype turns out to come from the photodetection noise, limited phase encoding resolution, and thermal crosstalk between phase shifters. Nevertheless, all these error sources can be relieved by different strategies in future implementations.

The above nanophotonic DNN still relies on regular computers at the training phase. This makes the whole process inefficient. An on-chip training scheme based on forward inference was proposed, which in principle could fit complex network architectures such as CNNs and recurrent neural networks (RNNs) where the effective number of parameters is substantially more than the distinct number of parameters. However, conducting such training requires repeatedly tuning every MZI in the circuit and is not efficient when the chip is scaled up. Hughes *et al.* developed an alternative protocol in which the on-chip training is accomplished only by *in situ* intensity measurements [168]. When a DNN is implemented by a nanophotonic platform employing the architecture in Ref. [23], the gradient terms of the cost function computed from backpropagation physically correspond to the error derivatives with respect to the permittivity of the phase shifters in the OIUs. Interestingly, this gradient distribution can be expressed as the solution to an electromagnetic adjoint problem. With the assistance of the adjoint variable method (AVM), an optimization technique used in the inverse design of photonics [169,170], the gradient at a phase shifter is given by the overlap of the optical fields from the "original" and "adjoint" problems. Figure 13(D) summarizes the training procedure demonstrated by simulations. The same flow applies if a real circuit is fabricated for experiments. The circuit has 3 input ports and 3 output ports to perform a 3×3 unitary matrix multiplication (panel (i)). Training starts by sending an (original) input vector $X$ from the left, e.g. $[0\ 0\ 1]^T$ as in panel (ii). The intensities of the light field at each phase shifter is measured and stored as $I_{og}$. Next, an adjoint field $\delta$, e.g. $[0\ 1\ 0]^T$ in panel (iii), is fed into the circuit from the output ports on the right. The field intensities are also measured and stored as $I_{aj}$. Based on the resulting field pattern, the time-reversed adjoint field $X_{TR}$ is calculated, which is then added to the original input $X$ to feed the unit from the input ports, resulting an interference pattern in panel (iv) and intensities $I$ at each phase shifter. The final computation of the gradient is done simply by subtracting $I_{og}$ and $I_{aj}$ from $I$, followed by a multiplication by a constant. The bottom two panels show the comparison of the gradient information, which are obtained by AVM with simultaneous excitations at both sides (panel (v)) and by the optical method (panel (vi)), namely interfering the patterns in panels (ii) and (iv), respectively. The very good agreement confirms that the gradient terms can be determined by *in situ* intensity measurements at the phase shifters. This result is significant, because it allows the computation in parallel of the crucial gradient distribution. With two OIUs in Figure 13(D) connected in series, a logic XOR gate was demonstrated. Figure 13(E) reports the network predictions before and after training, where the latter shows obvious improvement, matching perfectly with the truth table.

The above integrated nanophotonic neural networks are composed of hundreds of components to implement their functionalities. When the platform is scaled up to involve thousands or even millions of artificial neurons, the complexity of the achievable tasks can be dramatically boosted. Taking advantage of the analogy between information transformation through DNNs and light diffraction in layered structures, in a recent work, Lin *et al*. experimentally demonstrated various complex functions with an all-optical diffractive deep neural network ($D^2NN$) [171]. The mechanism is illustrated in Figure 14(A). Recalling the diagram in Figure 7, in a DNN, data are transformed from neuron to neuron via their interconnections. In a layered diffractive structure, according to the Huygens-Fresnel principle, each single point on a certain layer acts as a secondary light source. The individual points are excited by the incoming light waves from points in the preceding layer and emit light to the subsequent layer, resulting in, in theory, full "connectivity" between adjacent layers via diffraction and interference. Despite similarities in the layered structure and ways of connection, there are several differences between $D^2NNs$ and conventional DNNs. In standard DNNs, real-valued weights are associated with neuron connections, the neurons carry a nonlinear activation and an additive bias term, and the output of each neuron is the weighted sum of the real-valued inputs, computed by matrix multiplications. In contrast, $D^2NNs$ are complex-valued due to the nature of the optical waves. The output of each point is given by the product of the input wave and the complex-valued transmission or reflection coefficient at that point. In this process, the local transmission or reflection coefficient applies a multiplicative bias to the output wave at that point, which is then weighted through propagation to interfere with other secondary waves at points on the next layer, physically implementing the matrix multiplications. Two functions, namely a classifier (Figure 14(B)) and an imaging lens (Figure 14(C)), were experimentally demonstrated at 0.4 THz using 3D-printed D$^2$NNs with phase-only modulation, while the training process was still completed on a computer. The complexity of the tasks significantly increases the required number of neurons and of the training data. For the digit classifier, 55000 images of handwritten digits (0-9) were used to train a five-layer D$^2$NN composed of ~0.2 million neurons. Because of their more abstract feature representation than digits, fashion products could be daunting for classification. But remarkably, the numerical tests of a five-layer network showed a classification accuracy of 81.13%, and the experiments reached a 90% match with this result. Figure 14(D) shows a representative example of a sandal input image and the corresponding output intensity map (left column). As expected, most energy is directed to detector 5. The confusion matrix and energy distribution for the whole experimental test dataset can be found in the right column. Although the reported classification accuracy is lower than the record of 96.7% from state-of-the-art CNN algorithms [172], this D$^2$NN uses almost one order of magnitude fewer neurons and its performance can be improved by introducing amplitude modulation, additional layers, and possibly optical nonlinearity.

Lastly, DNNs are sometimes described as "black boxes", because it is almost impossible to extract an intuitive picture to explain how data are processed through the hidden layers. D$^2$NNs may provide some insights into this. In Figure 14(E), three spatially separated Dirac-delta functions are fed into a D$^2$NN-based unit-magnification imager composed of 10 layers of phase-modulation masks (left panels). It can be seen from the amplitude and phase distributions that indeed each neuron is connected to various neurons in the next layer in an abstract way. The input neither propagates as needle-sharp beams nor diffracts as in the free-space (right panel). Rather, each delta function tends to be diffused at the beginning, attenuated in the halfway, and finally focused to the same point of the output plane as where it is emitted on the input plane. Visualizing wave propagation through the D$^2$NN for this specific application help to reveal the operation principle of the coherent optical DNNs.

## 5. Conclusions and Outlook

In this review, we have summarized the recent advances on nanophotonics that are enabled or powered by advanced computational methods, especially deep learning algorithms. In the inverse design of nanophotonic devices, these techniques allow us to go beyond physical insights and help to search the parameter space in a more efficient way, leading to data-driven, on-demand design of novel devices. In the opposite direction, the development of nanophotonics could provide new platforms that can potentially overcome the bottleneck in computing power for machine learning. As the research interests and efforts on this topic continue increasing, we envisage that the following directions will be promising in the next stage of development.

First, advanced optimization techniques, especially gradient-based topology optimization that can handle up to 1 billion design variables efficiently [73,173], allow for the design of nanophotonic devices with tremendous complexity. Whereas this capability has been applied to the aerospace industry, the design resolution for nanophotonics is much lower. Increasing the number of design variables will empower the invention of more sophisticated and more integrated metadevices.

Second, the current application of deep learning in nanophotonic inverse design is still limited to finding appropriate design parameters for the desired spectra. Using low-dimensional inputs such as spectra and diffraction efficiencies [174] in the network ensures the data volume will not diverge but also restricts the achievable functionalities. Higher dimensional data such as the field intensity profiles as well as vectorial field maps carry much more information that can be used for designing functional metalenses and holograms. Blooming of nanophotonic devices enabled by deep learning is expected once the difficulties in computation power and data storage are overcome.

Third, the mechanism of many optical effects and multiphysics processes involving optics has not been well understood, which hinders the physics-inspired design but is where machine learning can come into play. For instance, although DNNs can provide accurate solutions when an arbitrary CD spectrum is desired, engineering the near-field optical chirality arising from the complex interplay between electric and magnetic fields is a task far from being solved. Similarly, optimizing optical forces at the nanoscale is critical for optical tweezers [114-117,175] and sorting [176-178]; nonlinearities of nanostructures for efficient harmonic generation and optical switching also have plenty of room to improve towards functional circuitry [179]. The lack of design guidelines makes them suitable problems for data-driven methods to deal with.

Next, despite the recent success in implementing DNNs on nanophotonic circuits and THz platforms, the all-optical realization of DNNs has not been fully demonstrated. For example, the training process is mostly conducted on electronic chips or computers, which does not really fulfill the advantages over speed and energy consumption. On-chip training, as numerically demonstrated, can overcome this limitation, while whether the losses will diminish the performance in a real system, and, how the measurement-based training can be effectively performed in large scale networks containing at least thousands of neurons, are potential issues to be addressed. $D^2NNs$ offer an alternative mechanism and platform to photonic circuits, whereas scaling down the THz scheme to visible or telecom wavelengths is demanding due to the limited fabrication resolution, interparticle coupling, and material losses. In addition, although demonstrated elsewhere, optical nonlinearity has not been introduced in photonic DNNs. Therefore, resolving these challenges will be essential steps towards all-optical DNNs [180].

Lastly, the use of machine learning techniques in nanophotonics has just emerged. Among the early attempts introduced in this review, many of them use standard network models, which may not be the best fit for the target applications. It is possible that the demonstrated performance can be simply improved by reforming the feeding data or modifying the network structure. Recurrent neural networks (RNNs) that can learn from sequential inputs have been realized on a photonic platform very recently but not yet in nanophotonics [181,182]. Including time domain features in the dataset will be very attractive. Furthermore, other leaning paradigms, such as unsupervised learning and reinforcement learning, and combinations of deep learning and other computational methods are expected to provide new design frameworks that are faster, more accurate, and even independent of human knowledge [183].

Nanophotonics and machine learning are two research domains that differ from the very basis. While it is promising to apply machine learning methods to data-driven nanophotonic design and discovery, many of the techniques, mature or cutting-edge, are not well known by the photonics community. Therefore, bridging this knowledge gap is pressing. Significant advancement will come out with further combination of the two fields.

**Acknowledgements**

The authors would like to acknowledge the financial support from the National Aeronautics and Space Administration (NASA) Early Career Faculty Award (80NSSC17K0520) and the National Institute of General Medical Sciences of the National Institutes of Health (DP2GM128446).

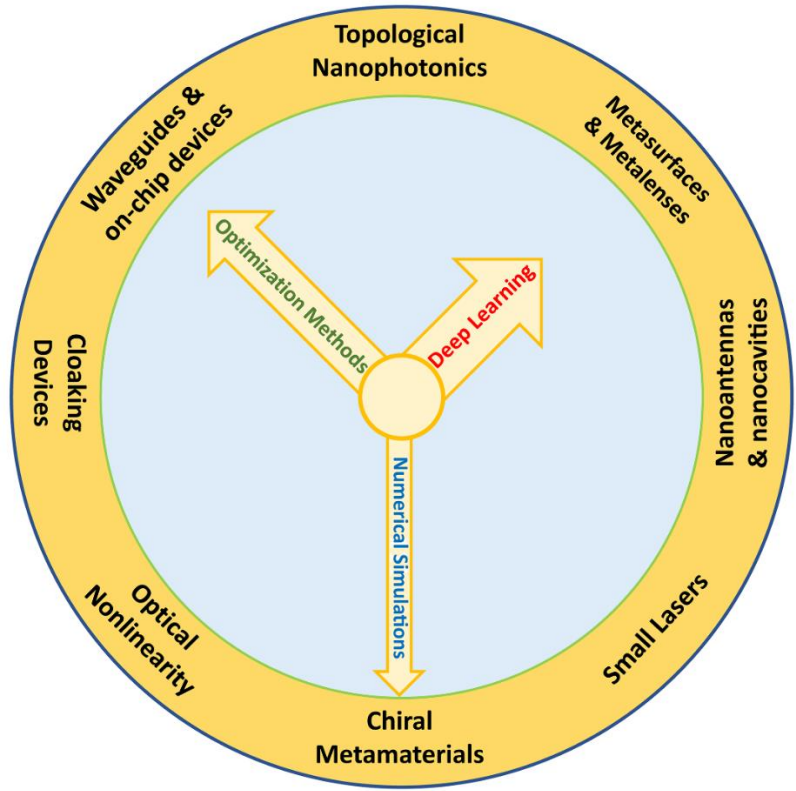

**Figure 1.** A dial illustration of computational methods and their potential applications in nanophotonics. Items on the circumference indicate different applications (not in the order of timeline or importance), where computational methods can be employed in the design process. The second hand corresponds to traditional schemes based on numerical simulations. Many cycles of trial-and-error modeling are needed for a specific task, each giving an incremental advancement towards the final goal. The minute hand denotes various optimization techniques, such as genetic/revolutionary algorithms and gradient-based approaches. Supported by simulation techniques, optimization methods search the full parameter space for each task by minimizing the cost function, providing a more efficient framework for achieving complex functionalities. The hour hand represents deep learning. Although a sufficiently large amount of data needs to be generated (by simulations) first for training, once the training is complete, the network solves a design request almost instantaneously.

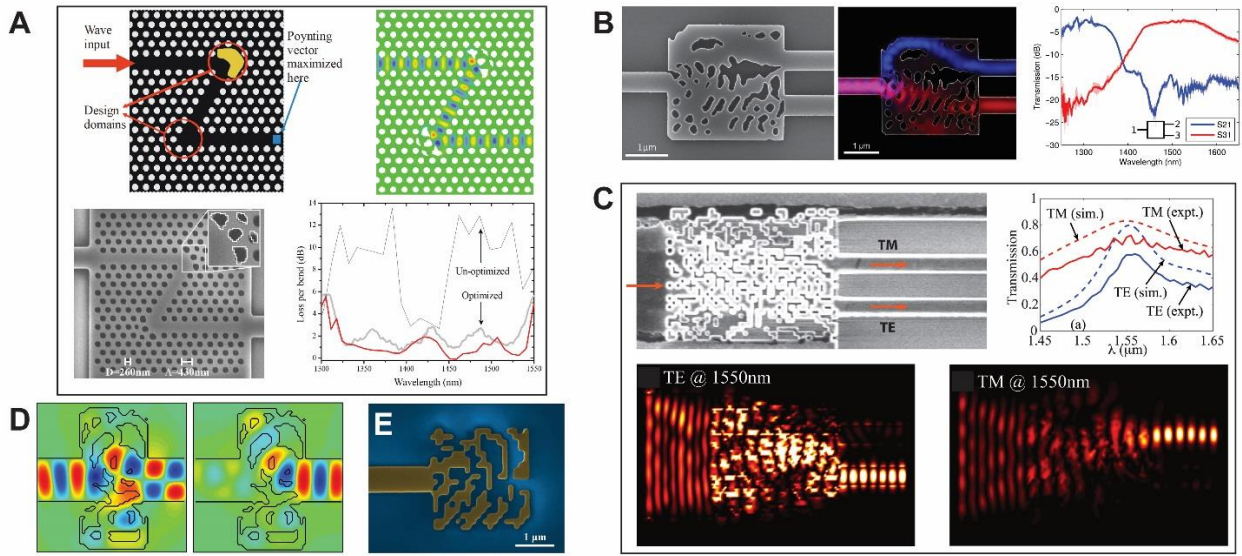

**Figure 2.** Demonstrations of on-chip devices based on inverse design. (A) A photonic crystal waveguide Z-bend showing exceptional transmission. Panels in zigzag order: Schematic of optimization; simulated wave propagation; SEM image of the fabricated Z-bend; comparison of bend losses between optimized (thick grey/red for measurement/simulation) and unoptimized (thin black) structures. (B) A two-channel wavelength splitter. Left: SEM image. Middle: Simulated field patterns for 1300 (blue) and 1550 nm (red) wavelengths. Fields are superimposed and color-coded for illustration. Right: Measured transmission spectra. (C) A nanophotonic mode-converting polarization beam splitter. Top: SEM image and comparison of measured (solid lines) and simulated (dashed lines) transmission of the fabricated device. Bottom: Simulated field intensity distributions for TE and TM modes at 1550 nm. (D) Simulated magnetic field distributions in an optical diode when the excitation is on the left (left panel) and right (right panel), respectively. (E) SEM image of a photonic reflector, as part of an on-chip Fabry-Pérot cavity. (A) is reprinted with permission from Ref. [74], OSA; (B) is reprinted from Refs. [11] and [76] by permission from Springer Nature; (C) is adapted from Ref. [80] by permission from Springer Nature; (D) is reprinted from Ref. [71] with permission (CC BY 4.0); (E) is reprinted from [82] with permission.

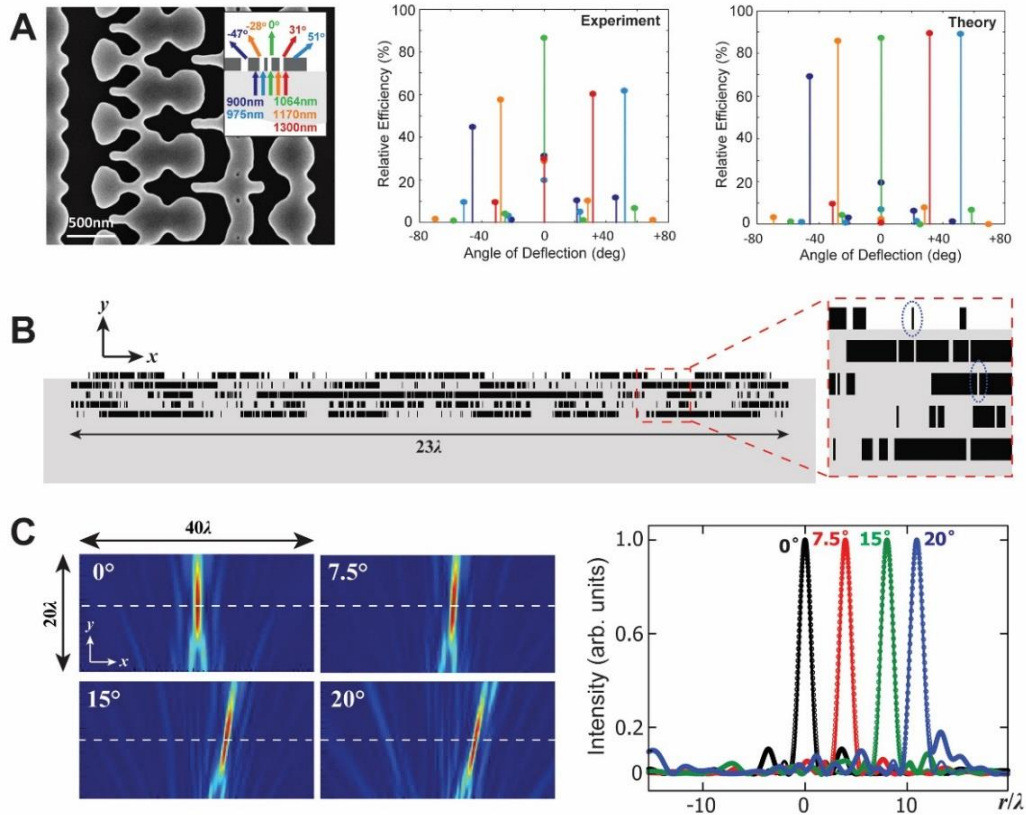

**Figure 3.** Metasurface inverse design using topology optimization. (A) A five-wavelength beam splitter with the SEM image (left), experimental (middle) and theoretical (right) diffraction efficiencies. Inset: Correlation between incidence wavelengths and deflection angles. (B) Design of a multilayer focusing metalens with angular aberration correction. Rectangles in black denote silicon resonators, and the grey background is alumina. (C) Simulated far-field intensity profiles for the structure in (B) at four angles of incidence follow the identical diffraction limit. Intensities are normalized to unity for comparison. (A) is reprinted with permission from Ref. [83], Copyright 2018 John Wiley and Sons; (B) and (C) are adapted with permission from Ref. [89], Copyright 2018 American Physical Society.

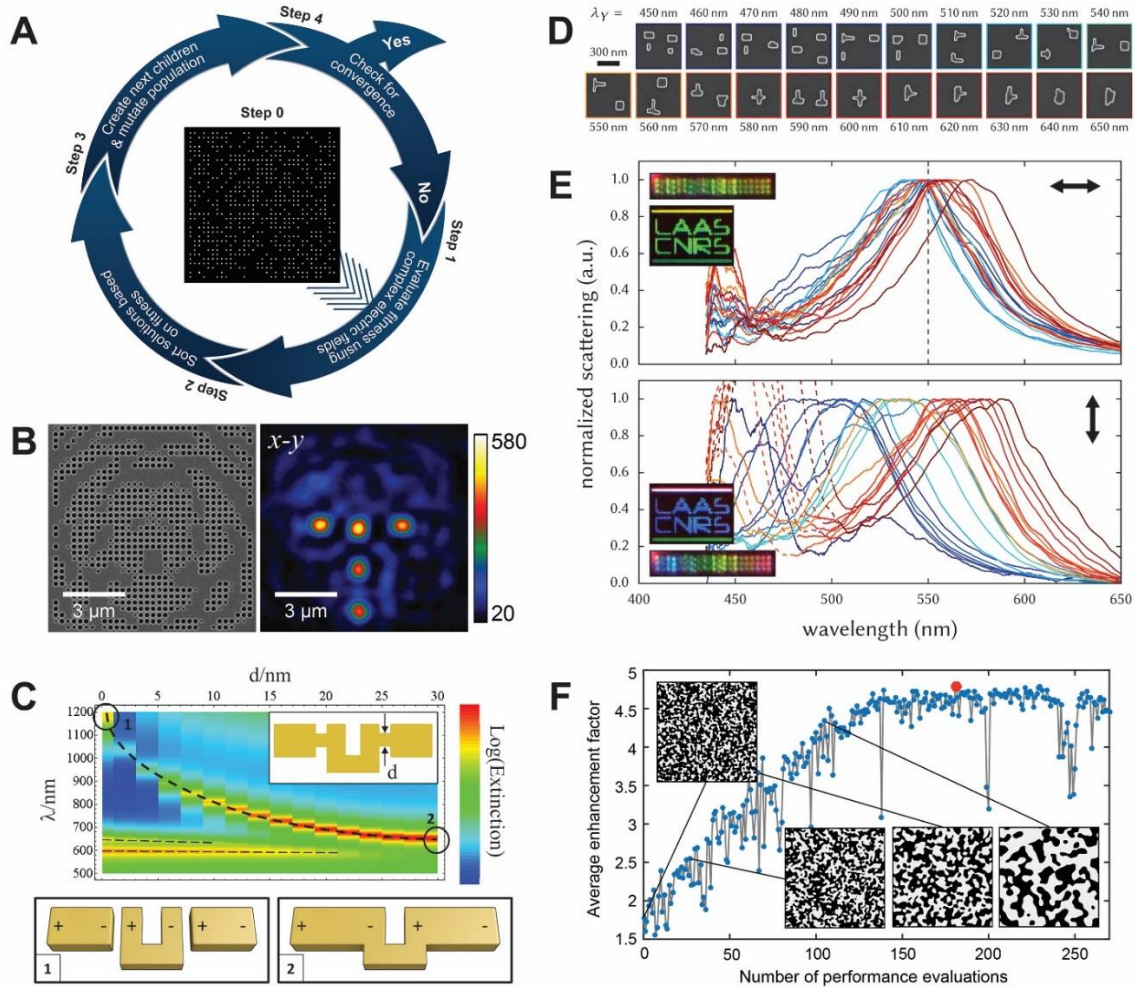

**Figure 4.** Nanophotonic devices based on evolutionary design and optimization. (A) Illustration of procedures of evolutionary design. (B) The SEM image (left) of a metasurface that can generate five focal points (right) arranged in a T-shape. (C) A simplified model of optimized matrix nanoantennas for improving the near-field intensity enhancement. (D) SEM images of optimized silicon nanoantennas for polarization-encoded color display. The associated wavelengths denote the resonances for *y*-polarized incidence, while for *x*-polarization, the resonance is targeted at 550 nm. (E) Measured scattering spectra for *x*- (top) and *y*-polarized (bottom) incidence. (F) Optimization history of the averaged absorption enhancement in a quasi-random light-trapping structure. Insets show the generated patterns at selected optimization steps. (A) and (B) are reprinted from Ref. [90] with permission; (C) is reprinted from Ref. [92] with permission, Copyright 2018 American Physical Society; (D) and (E) are adapted from Ref. [94] by permission from Springer Nature; (F) is reprinted from Ref. [95] with permission, Copyright 2018 National Academy of Sciences.

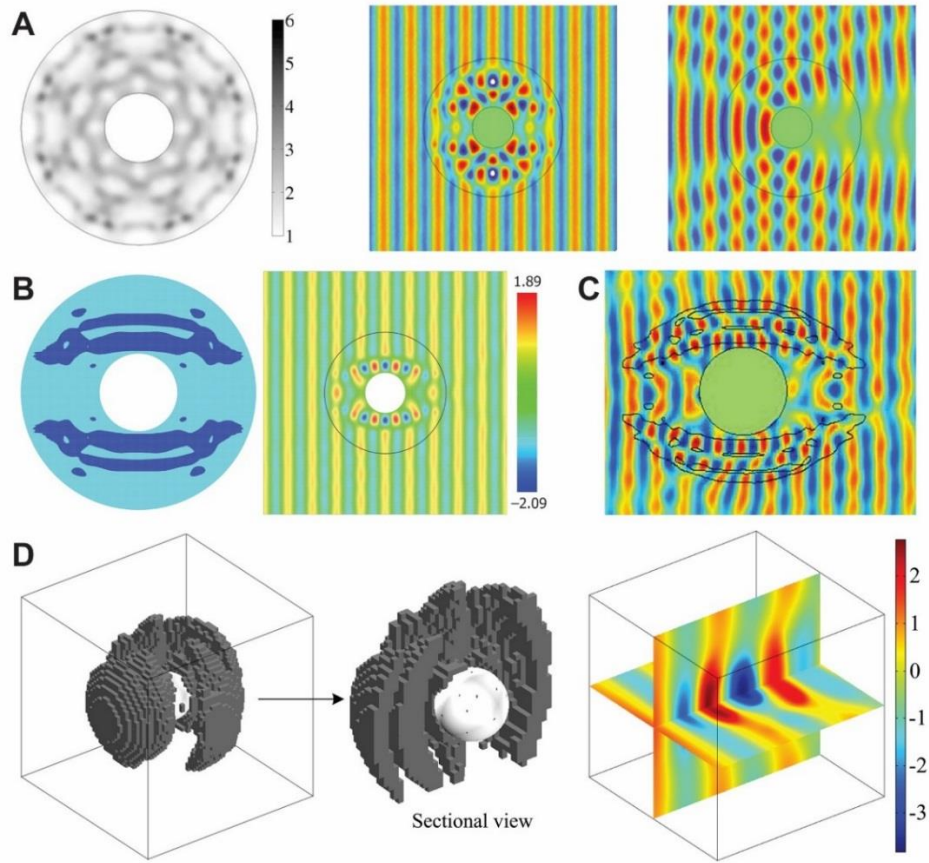

**Figure 5.** Topology-optimized cloaks and concentrators. (A) A 2D low-contrast all-dielectric cloak. Left: The dielectric layout for four symmetry lines. Inner circle denotes an ideal metallic cylinder. Middle: Numerical demonstration of the cloaking performance. Right: The scattering pattern of a bare cylinder. (B) A 2D unidirectional cloak with a low refractive index material (blue region, $\varepsilon_r = 2$) and the simulated scattering pattern. (C) Experimental realization of the design in (B) at microwaves. (D) Design of a 3D magnetic field concentrator (left) and the simulated magnetic field distribution (right). (A) and (B) are adapted from Refs. [106] and [107], respectively, with permission of AIP; (C) is reprinted from Ref. [108] with permission of AIP; (D) is reprinted from Ref. [110] with permission.

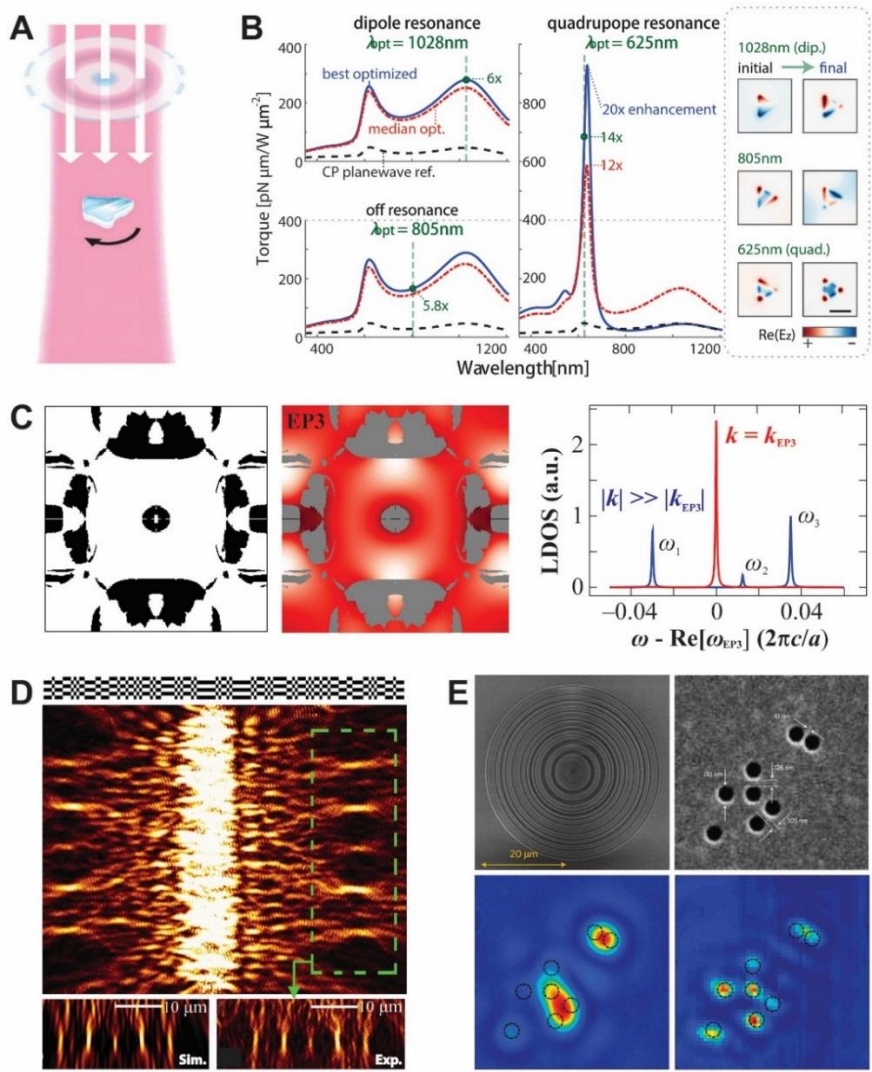

**Figure 6.** (A) Inverse design of illumination patterns to maximize optical torques. (B) Optimized torque spectra and field distributions for three target wavelengths at 1028, 805, and 625 nm, respectively. (C) Inverse design of PhCs for enhancing spontaneous emission. Left: Dielectric layout of the design. Black regions correspond to a material with refractive index $n = 2$. Middle: LDOS profile at the third-order Dirac exceptional point. Right: LDOS spectra at the center of the unit cell. (D) A binary plasmonic structure (top) designed by simulated annealing algorithms to produce five focal points of SPPs. (E) A super-oscillatory lens designed by particle swarm optimization for subwavelength imaging. Top: SEM images of the fabricated lens (left) and a cluster of nanoholes in a metal film (right). Bottom: Images of the cluster by a conventional lens (left) and by the super-oscillatory lens (right). (A) and (B) are reprinted with permission from Ref. [120], OSA; (C) is adapted with permission from Ref. [121], Copyright 2018 American Physical Society; (D) is adapted with permission from Ref. [123], OSA; (E) is adapted from Ref. [124] by permission from Springer Nature.

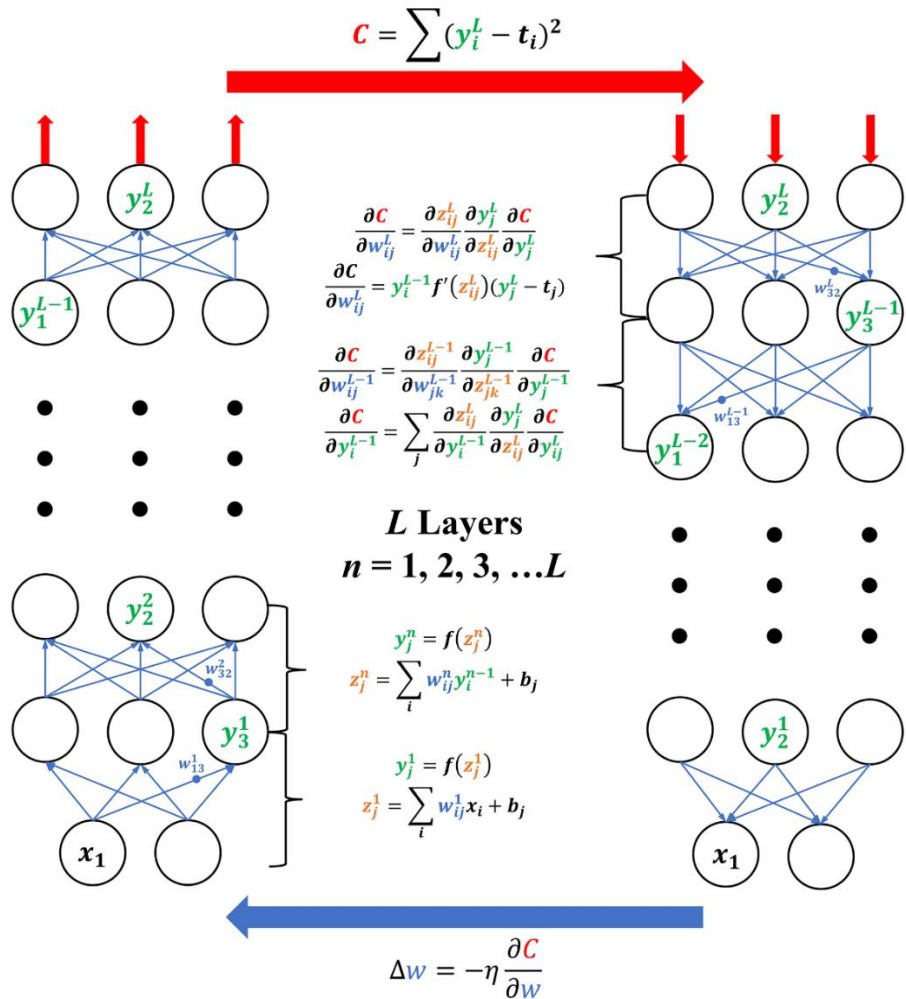

**Figure 7.** The architecture and learning process of a deep neural network comprising multiple layers. The circles represent artificial neurons. In the shown case, neurons in the same layer do not interact, but each neuron is connected, with a unique weight $w_{ij}$, to every neuron in the adjacent layer(s), namely the preceding and/or subsequent layers. Data transformation between adjacent layers is implemented by linear matrix multiplications of the weights and input vectors, followed by the application of a nonlinear function $f$ (and sometimes an additive bias $b_j$) at each neuron. Left: Forward inference in which data flow through the network from the input layer to the output layer. Right: Training with backpropagation, where every weight value is adjusted based on the error derivative to minimize the cost function. The forward inference, backpropagation, and weight update are repeatedly performed as data are continuously supplied, until the desired performance is obtained.

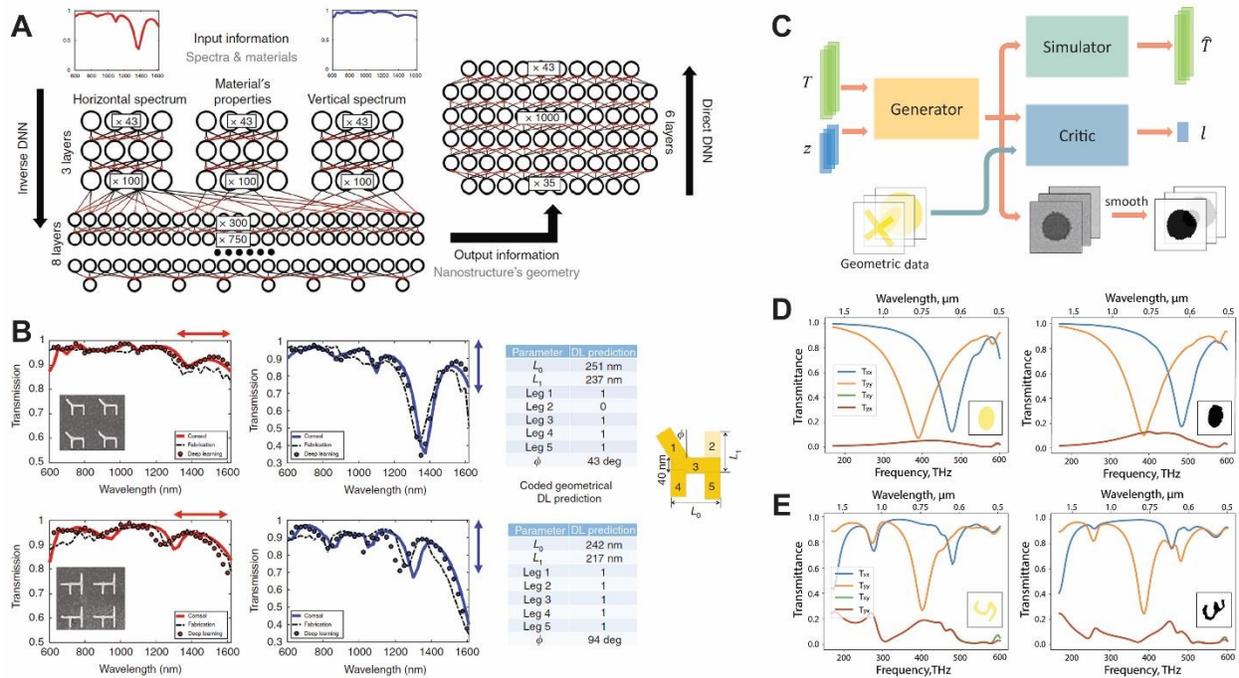

**Figure 8.** (A) A DNN for design and characterization of metasurfaces. The network comprises a layered GPN (left) to solve the inverse design problem and an SPN (right) to predict the spectra based on retrieved design parameters. (B) Demonstration of design retrieval and spectra prediction based on (A). The building block is a gold nanostructure with its geometry represented by a general "H" form. (C) Architecture of a generative network composed of a generator, a pre-trained simulator, and a critic. In the training phase, by receiving spectra $T$ and the corresponding patterns $X$ respectively, the generator and the critic learn jointly. Valid patterns are smoothened to binary maps and stored as candidates for metasurface design. (D) Original (left) and generated pattern (right) and their transmittance spectra based on a training dataset of elliptical nanoparticles. (E) Original (left) and retrieved pattern (right) and their transmittance spectra based on an incomplete dataset of handwritten digits. The network generated a modified "3" to best replicate the spectra for pattern "5", which was removed intentionally from the training data. (A) and (B) are reprinted from Ref. [20] with permission (CC BY 4.0); (C)-(E) are reprinted with permission from Ref. [22], Copyright 2018 American Chemical Society.

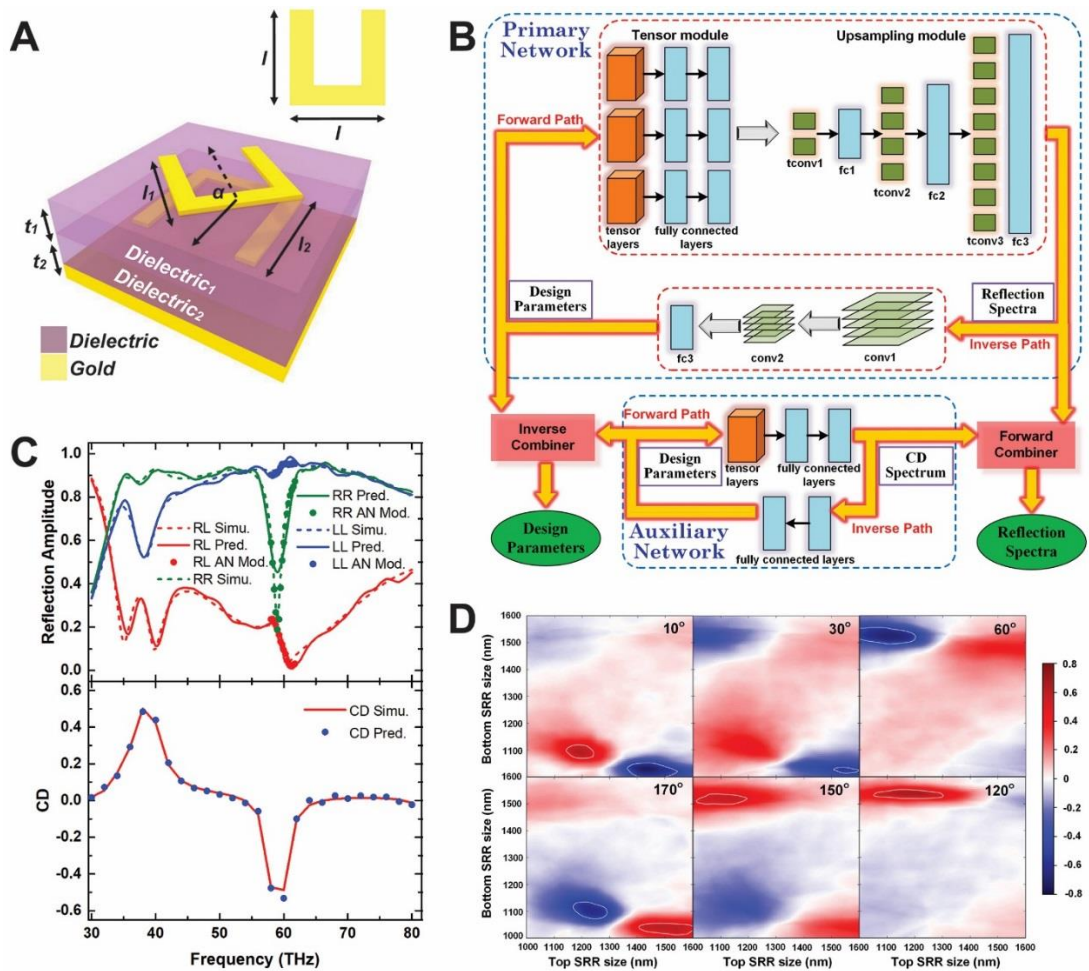

**Figure 9.** (A) The unit cell of a chiral metamaterial consisting of two layers of gold SRRs on top of a reflective mirror. Two SRRs are twisted by an angle $α$, which breaks the mirror symmetry of the system and determines the chiral response together with other design parameters, including the sizes of SRRs and the thicknesses of the spacing layers. (B) A combined architecture consisting of two bidirectional DNNs. The primary network (top) connects design parameters and reflection spectra, and the auxiliary network (bottom) creates mappings between design parameters and CD with higher accuracy. Data are allowed to flow between PN and AN via combiners to refine resonance features in all the spectra. (C) Top: Comparison of reflection spectra obtained by simulations (dashed lines), by PN alone (solid lines), and by the combined system in (B) (dotted lines). Bottom: Comparison of simulated CD and AN predicted CD. (D) Evolution of CD at 60 THz for different combinations of SRR sizes and selected twisting angles. Figures are adapted with permission from Ref. [21]. Copyright 2018 American Chemical Society.

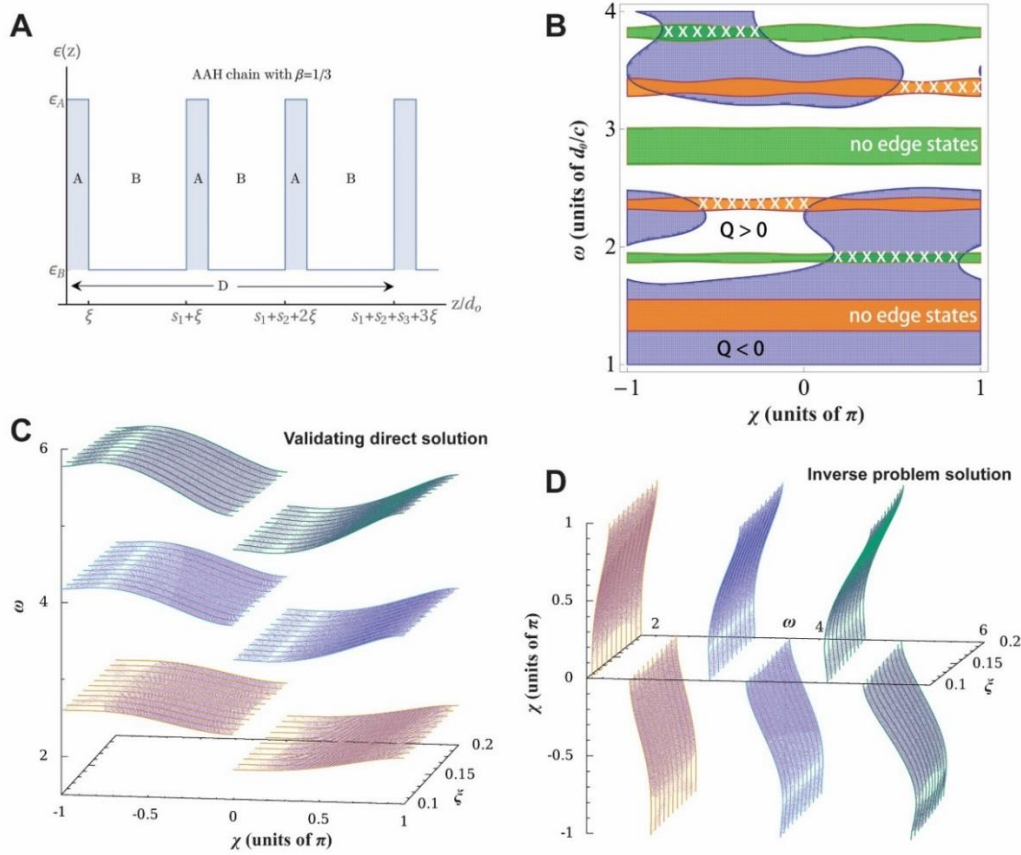

**Figure 10.** (A) Dielectric function profile of a multiplayer structure with Harper modulation. Layers with material A are stacked with spatial modulation along the *z*-axis in a homogeneous bulk material B. (B) The band diagram of the structure in (A). Orange and green ribbons represent bandgaps. Edge states are denoted by regions with crosses, which exist only in the bandgaps where a complex quantity Q changes sign. White and purple regions correspond to $Q > 0$ and $Q < 0$, respectively. (C) The band diagram predicted by the direct DNN. (D) The modulation diagram retrieved by the inverse DNN. In (C) and (D), colored curves represent the training dataset. Figures are adapted from Ref. [148] with permission (CC BY 4.0).

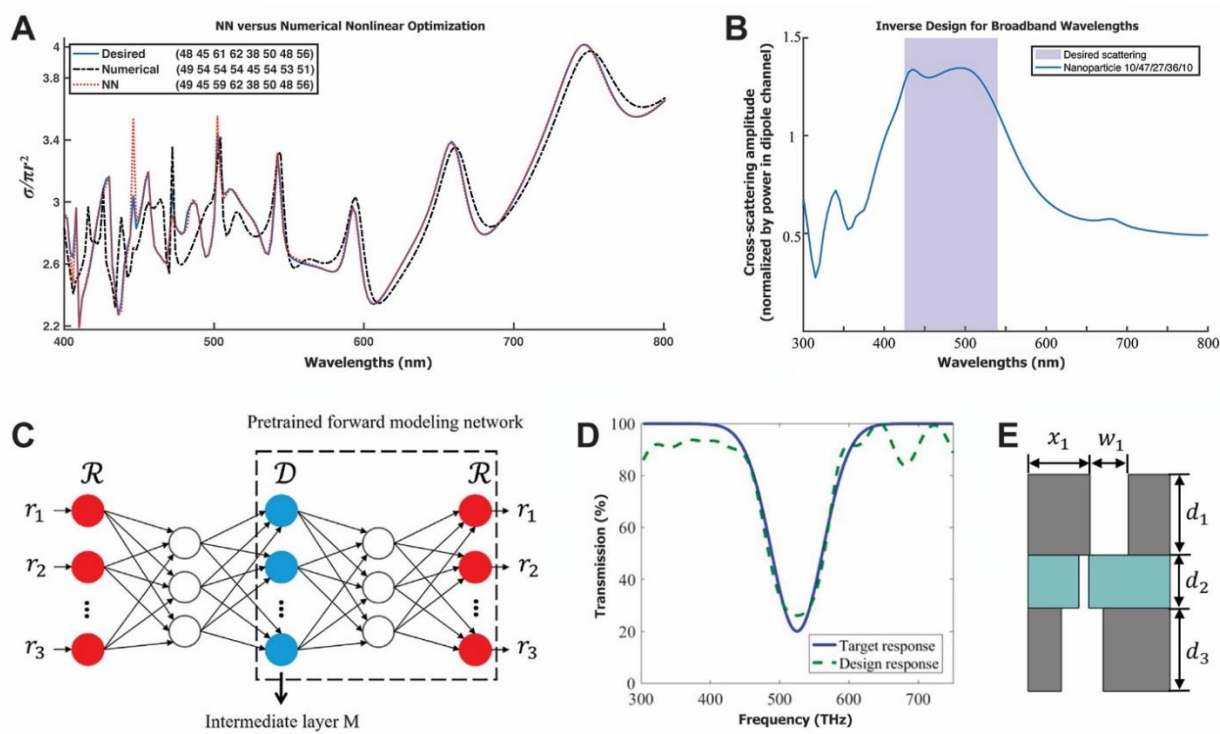

**Figure 11.** Inverse design of multilayer structures via deep learning. (A) A DNN retrieves the layer thicknesses of a multilayer particle based on its scattering spectrum, showing much higher accuracy than the nonlinear optimization method. The main figure compares the scattering spectra by simulation (blue), optimization (black), and prediction of DNN (red). Comparison between the ground truth and retrieved design parameters are given in the legend. (B) Inverse design of multilayer particles for broadband scattering in a given wavelength range. (C) A tandem network for designing multilayer thin films. A pretrained forward modeling network is connected to the inverse design network to avoid non-uniqueness issues. (D) Performance of the tandem network in fitting a Gaussian profile in the frequency domain. (E) Structured multiplayer thin films for modulating the transmission phase delay. (A) and (B) are reprinted from Ref. [151] with permission (CC BY-NC); (C)-(E) are reprinted with permission from Ref. [152], Copyright 2018 American Chemical Society.

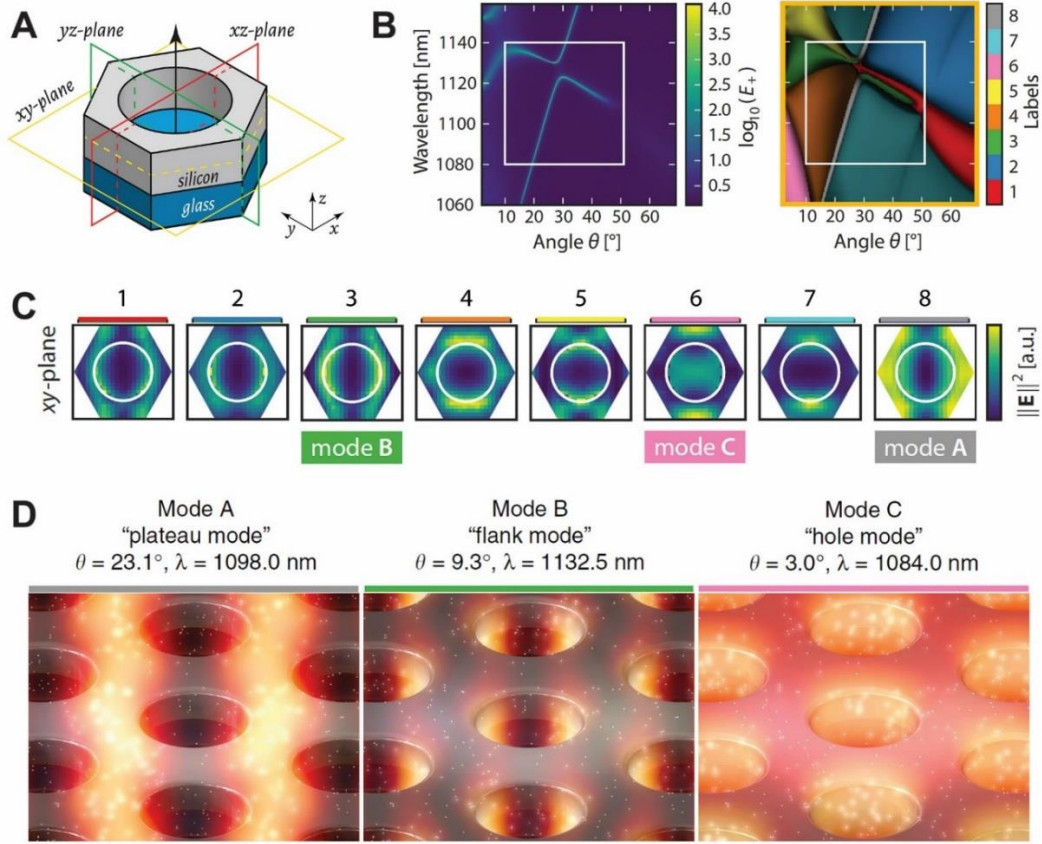

**Figure 12.** (A) The unit cell of a silicon PhC on glass, in which circular holes form a hexagonal lattice. When illuminated by external light from the top, leaky modes are excited, which exhibit strong near-fields boosting the emission of the nearby emitters, such as quantum dots. Rectangles in color denote the symmetry planes used for exporting field data. (B) Comparison of the simulated band diagram (left) and clustering results (right) when the sample is oriented in the Γ-K configuration and irradiated by a TE-polarized plane wave. Left: The diagram of volume-averaged electric field energy enhancement. Right: The classification map depicted by blending the color-coded silhouette coefficient of each mode with a black background. (C) Top view of electric field energy plots for the eight prototypes found in the clustering, with three of them associated with leaky modes exhibiting strong near-fields. Mode A: plateau mode; Mode B: flank mode; Mode C: hole mode. Modes are termed base on the location of the field enhancements. (D) 3D semi-artistic plots of the interactions between leaky modes and randomly distributed quantum dots, which emit light with an intensity proportional to local field energy density. Figures are adapted from Ref. [153] with permission (CC BY 4.0).

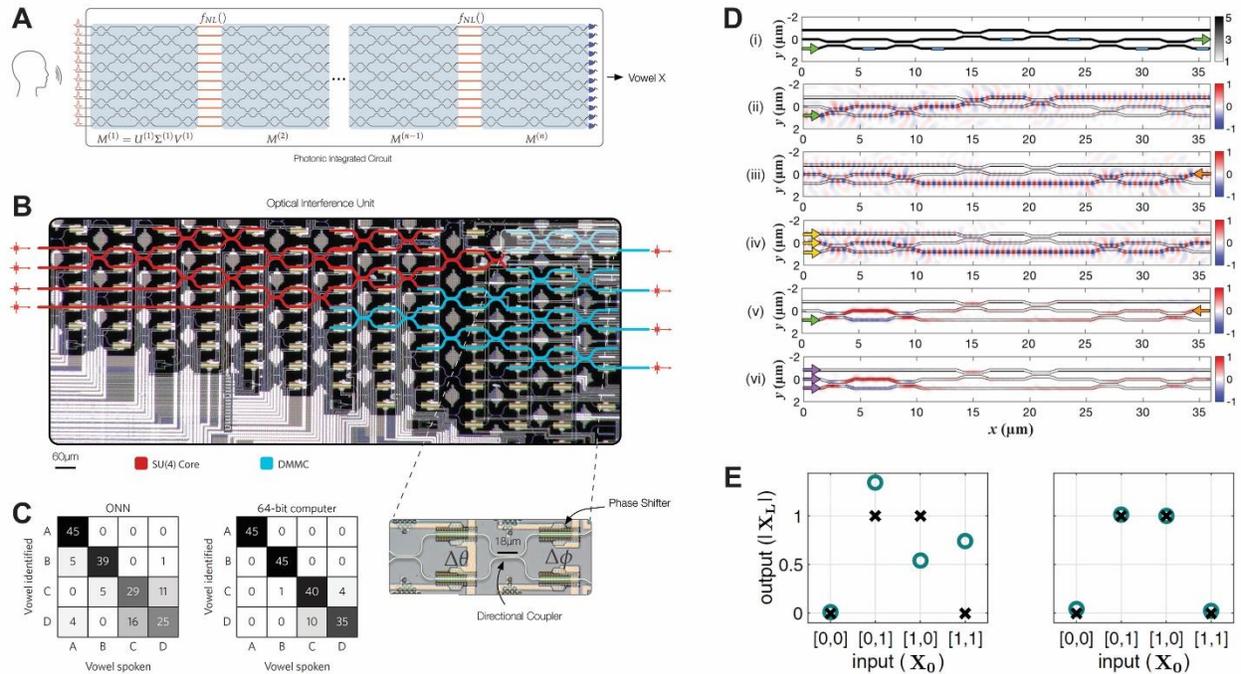

**Figure 13.** Nanophotonic DNNs. (A) The architecture of a nanophotonic neural network for vowel recognition. Each box in grey corresponds to an optical interference unit that computes matrix multiplication, followed by an optical nonlinearity unit connecting it to the next layer. (B) Optical micrograph of the optical interference unit used in the experiments. The unit comprises 56 programmable MZIs, with the red/blue mesh highlighting the functional part implementing a multiplication by a 4×4 unitary/diagonal matrix. Inset: Layout of the MZIs composed of two phase shifters and a directional coupler. (C) Confusion matrix of the nanophotonic circuit (left), in comparison to that of a 64-bit computer (right), for vowel recognition by a network with two hidden layers. The elements (X,Y) of the matrices are the numbers of times a spoken vowel X is identified as Y. Perfect identification would give a diagonal matrix. (D) On-chip training of a nanophotonic OIU through *in situ* measurements. (E) An optically trained nanophotonic DNN implementing a logic XOR gate. The tables compare the network predictions before (left) and after (right) training. Target answers (truth table) are depicted with crosses and predictions are denoted by circles. (A)-(C) are adapted from Ref. [23] by permission from Springer Nature; (D) and (E) are adapted with permission from Ref. [168], OSA.

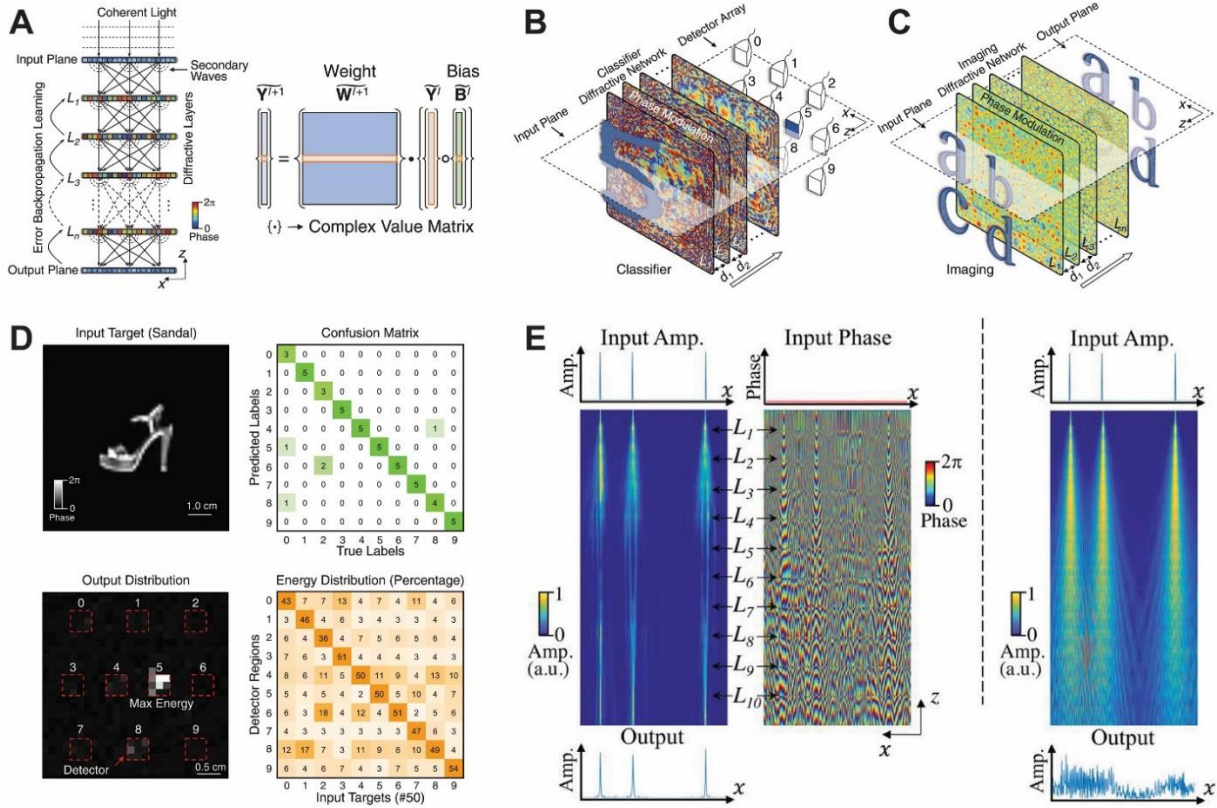

**Figure 14.** All-optical machine learning by D$^2$NNs. (A) A D$^2$NN consisting of multiple transmissive layers. Each point (color coded squares) acts as a neuron by providing a multiplicative complex-valued transmission coefficient to the incoming wave. The coefficient distribution is trained by deep learning algorithms to perform a predefined function and is fixed after fabrication. (B) Schematic of a 3D-prined N$^2$DD implementing a classifier for handwritten digits and fashion products. Different types of objects on the input plane leads to maximized light intensity at the corresponding detector on the output plane. (C) Schematic of a 3D-printed N$^2$DD implementing a lens for imaging. The output is a unit-magnification image of the input optical field. (D) Classification of fashion products. Left: An input image belonging to a certain class of products, e.g. sandals, results in maximum energy at the corresponding detector on the output plane. Right: Confusion matrix and energy distribution for the full test dataset of the fashion products classifier. (E) Wave propagation in a N$^2$DD imager with 10 layers. Left: Amplitude and phase distributions for an input of three Dirac-delta functions passing through the network. Right: Amplitude distribution for the same input passing through a vacuum. Diagrams at the bottom show the light intensities on the output plane. Figures are reprinted from Ref. [171] with permission from AAAS.